\documentclass[12pt]{article}

\usepackage[utf8]{inputenc}
\usepackage{graphicx}
\usepackage{amsmath,amssymb}
\usepackage{placeins}          
\usepackage{float}             
\usepackage{subcaption}
\usepackage[labelfont=bf,skip=4pt]{caption}
\usepackage{geometry}
\usepackage{textcomp}          

\usepackage[numbers,sort&compress]{natbib}

\title{Generation of UV-Vis correlated photon pairs using PP-LaBGeO$_5$}
\author{Ricardo Tolentino, Nathan Harper, and Scott Cushing\\[0.3em]
\small Division of Chemistry and Chemical Engineering,\\
\small California Institute of Technology, Pasadena, CA 91125, USA\\
\small \texttt{scushing@caltech.edu}}
\date{} 

\newcommand{\authormark}[1]{}
\newcommand{\address}[1]{}
\newcommand{\email}[1]{}
\newenvironment{abstract*}{\begin{abstract}}{\end{abstract}}
\providecommand{\JournalTitle}[1]{#1}
\usepackage{hyperref}          
\begin{document}

\maketitle

\begin{abstract*} 
The SHG and SPDC output of a periodically poled bulk LBGO crystal is measured. The SHG efficiency is measured to be (3.52 +/- .04) x 10$^{-3}$ \%/W. The SPDC output is temporally characterized by a Hanbury-Brown-Twiss interferometer; a maximum CAR of 244 and pair generation rate of (1.45 +/- .47) x 10$^6$ $\mathrm{pairs\,s^{-1}\,mW^{-1}}$ is measured. An EMICCD spectrometer is used to measure the spectral characteristics of the PP-LBGO source. Of the four gratings tested, one operational grating (poling period = 2.10 \textmu m) of the PP-LBGO device outputs a temperature tunable spectrum of entangled photon pairs centered at 532 nm with signal and idler photons being tunable from 507 to 556 nm, respectively. The work paves the way for future waveguided entangled photon sources reaching UV wavelengths using periodic poling.

\end{abstract*}

\section{Introduction}
 Spontaneous parametric down-conversion (SPDC) has predominantly been utilized at the near-infrared (IR) and telecom wavelengths in quantum networks and microscopy. Many chemical and photochemical systems with conjugated pi bonds require ultraviolet (UV) to visible light to initiate electronic transitions, excited state dynamics, photochemical reactions, or photodegradation. SPDC sources where both photons are at visible wavelengths (350-700 nanometers (nm)) are required, necessitating a UV pump (<300 nm). In this wavelength range, both photons can be efficiently absorbed by fluorescent dye molecules and are readily measured with silicon detectors. This enables classical limits in spatial, spectral, and temporal resolution to be surpassed. For example, quantum entanglement-enabled super-resolution microscopy enables microscopy with resolution beyond the diffraction limit, allowing for enhanced spatial resolution and image contrast\cite{Camphausen2023,He2023,Giovannetti2004,Rozema2014}. Another exciting use case is time-resolved fluorescence (TRF) spectroscopy. Here, the entangled pair is utilized as two single-photon packets; one photon of the pair initiates a fluorescence event while the other is detected in coincidence\cite{Harper2023,Eshun2023}.
 In this case, a TRF measurement can be performed without requiring a pulsed laser. Finally, the spectrum from a single SPDC source is broadly tunable from the wavelength of the pump laser to the far-IR\cite{Wang2020}. Even classical lasers struggle to match this tunability, requiring ultrafast pulses and multiple nonlinear stages. 

Access to the sub-532 nm regime is challenging when selecting suitable photonic materials that can be periodically poled and waveguided. Traditionally, lithium niobate or lithium tantalate is selected. However, their strong absorption in the deep UV region dampens their effectiveness toward nonlinear processes in the deep UV to UV regime\cite{Wang2020, Harper2024a, Qi2020, Thyagarajan2009}. Very few nonlinear materials exist with transparency at 266 nm. Beta barium borate ($\beta$-BBO) has been used to generate photon pairs pumped at 266 nm. The birefringent phase matching configuration inherent to $\beta$-BBO results in low pair generation rates due to a weak nonlinear tensor element and limited interaction lengths due to spatial walk-off\cite{Eshun2023, Davenport2024, Brown2023, Lavoie2020}. These issues are resolved by using type-0 phase matching in quasi phase matched (QPM) devices. Periodically poled LaBGeO5 (PP-LBGO) is an interesting alternative with a transparency extending to 195 nm\cite{Hirohashi2014a, Hirohashi2014b, Hirohashi2015a, Hirohashi2015b, Miyazawa2016, Nakahara2018, Watanabe2019a, Hirohashi2019a, Hirohashi2019b, Umemura2019, Akhmatkhanov2020, Sakairi2020, Tanaka2021, Plashinnov2021, Shoji2022, Yamanobe2022}.

PP-LBGO is an emerging frequency conversion material of interest owing to its high UV transparency and non-hygroscopicity as a QPM device. Previous studies have investigated harmonic generation and reported the values of PP-LBGO’s second-order nonlinear optical coefficient. Due to the wavelength dependence of the nonlinear tensor, efforts have been made to determine and reconcile the coefficients at 1064 and 532 nm. The most recently reported nonlinear coefficient values measured at 532 nm are d$_{11}$ = 0.55, d$_{22}$ = 1.41, d$_{33}$ = 1.15, d$_{31}$ = d$_{32}$ = 0.75 pm/V\cite{Tanaka2021}. These coefficients are used to predict the Second Harmonic Generation (SHG) and SPDC properties of PP-LBGO with $d_{\text{eff}}$  =  0.73 pm/V for type-0 phase matching.

We measure SHG and SPDC in periodically poled LaBGeO5 (PP-LBGO). An SHG efficiency of (3.52 +/- .04) x 10$^{-3}$ \%/W (percent efficiency per Watt) and an SPDC efficiency of (1.45 +/- .47) x 10$^{6}$ pairs per second per milliwatt ($\mathrm{pairs\,s^{-1}\,mW^{-1}}$) is measured. Temporal characterization is obtained by a Hanbury-Brown-Twiss interferometer, and we report a max coincidence to accidental ratio, or CAR of 244. The SPDC output is centered at 532 nm and can be temperature tuned from 507 to 556 nm. The overall efficiency is low compared to waveguided periodically poled lithium niobate based devices in the NIR-VIS, motivating the need for exploration of waveguiding and similar methods in this emerging material if efficient UV-SPDC is to be achieved.

\section{Experimental Details}

The Sellmeier coefficients for LBGO predict that type-0 SPDC, where all three interacting fields have identical, extraordinary polarizations, is phase-matched for 266 nm SHG at ~66 Celsius (°C) with a first-order poling period of 2.1 \textmu m. A bulk PP-LBGO device was obtained from Oxide Corporation with poling periods around this value (2.1, 2.15, 2.20 and 2.25 \textmu m). The 2.10 \textmu m grating was experimentally found to be phase-matched for both SHG and SPDC and is used for the reported measurements. The device is 0.3 millimeters (mm) thick, each grating is 1 mm wide, and the crystal is 10 mm long. The chip is mounted to a temperature-controlled block on a three-axis translation stage. The chip is heated from room temperature to 80 °C with a stability of .1 °C. The Sellmeier equations used for design considerations are taken from the work of Umemura et. al \cite{Umemura2019}.

To characterize the SHG performance of our LBGO crystal, we utilize the following experimental setup (Figure 1). It consists of a single-frequency diode-pumped solid state (DPSS) laser source at 532 nm operating at a max power of .28 watts (W), which is passed through a bandpass filter and half-waveplate before being focused through the grating using a 10 centimeter (cm) focal length lens. The output, now consisting of the generated 266 nm second harmonic and the residual 532 nm pump beam, is collimated and passed through a filter stack consisting of 4 short-pass filters for a total optical density (OD) 20. The filtered 266 nm signal is then coupled into a multimode fiber, which is used as the input to either a spectrometer for spectral characterization or a UV-sensitive photodiode to measure the generated power. 

\begin{figure}[htbp]
\centering\includegraphics[width=13cm]{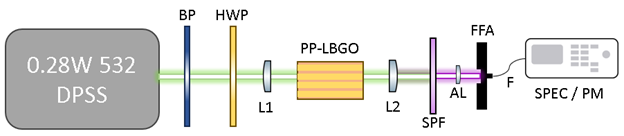}
\caption{SHG characterization setup of PP-LBGO. BP – Bandpass Filter, HWP – Half Wave Plate, L – Lens, SPF – Short Pass Filter, AL – Aspheric Lens, FFA – Free space to Fiber Adapter, F – Multimode Optical Fiber, SPEC – Spectrometer, PM – Power Meter.}
\end{figure}

\begin{figure}[htbp]
\centering\includegraphics[width=13cm]{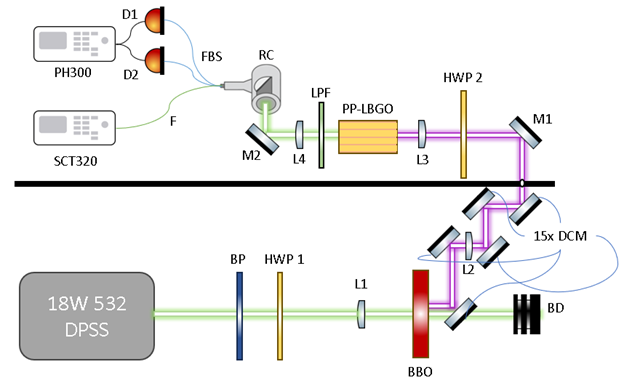}
\caption{UV pumped SPDC. BP – Bandpass Filter, HWP – Half Wave Plate, L – Lens, BD – Beam Dump, DCM – Dichroic Mirror, M – Mirror, LPF – Long Pass Filter, RC – Reflective Collimator, FBS – Fiber Beam Splitter, F – Optical Fiber, D1/D2 - SPAD, PH300 – Coincidence Counter, SCT320 - Spectrometer}
\end{figure}




A Coherent Verdi V-18 DPSS Continuous Wave (CW) Laser pumps the SPDC characterization setup (Figure 2). The laser outputs 100:1 s-polarized 532 nm light with a beam diameter of 2.25 mm. The beam passes through a bandpass filter and half waveplate before being focused into a BBO by a 100mm lens mounted on a linear stage and a two-axis stage for tip/tilt adjustment. We filter the pump laser through a series of 15 dichroic mirrors with a high reflectivity at 266 nm. The DCM mirrors reduce the residual pump to a maximum of 5x10$^{3}$ counts/s on the SPADs. A beam dump behind the first mirror is responsible for dissipating 95\% of the energy of the pump laser. A collimating lens (L2) is placed 100 mm after the BBO in the cascaded dichroic mirror setup. The 266 nm beam passes through a pinhole and is reflected by M1 toward a half waveplate before focusing through the PP-LBGO device. The PP-LBGO is mounted on a 5-axis stage equipped with a resistive heater. The output passes through a series of long pass filters before being collimated by a 75 mm lens (L4). A reflective collimator (RC) couples the SPDC into a multimode fiber connected to a Hanbury-Brown-Twiss (HBT) interferometer for the coincidence measurements (blue path), or directly to a spectrometer (green path). The HBT interferometer is comprised of a fiber beamsplitter (Thorlabs TM105R5F1A) connected to two identical SPAD devices (Hamamatsu C13001-01) which record photon coincidences on a coincidence counter (PicoHarp 300).The spectrometer consists of an IsoPlane SCT 320 with a grating (I3-015-800-P) and a camera (PI-MAX 4 1024i 50).  

\section{Results}

We performed a temperature scan of the crystal to characterize the temperature dependence of the phase-matching, as shown in Fig 3a. For the 2.1 \textmu m poling period of our crystal, we expected the phase-matching temperature to be 65.7 °C, as shown in Fig 3b. The experimentally measured value which resulted in the strongest 266 nm signal was found to be 65.2 °C. We attribute the phase-matching temperature discrepancy to the SHG wavelength being at the lower limit of the Sellmeier equations used in this analysis.

The SHG experimental setup is used to characterize the single-pass efficiency of the LBGO crystal. 2.74 microwatts (\textmu W) of the 266 nm second harmonic is generated when the device is pumped with 279 milliwatts (mW) of the 532 nm first harmonic. These values correspond to an SHG efficiency of 9.82 x 10$^{-4}$ \% at 279 mW, or (3.52 +/- .04) x 10$^{-3}$ \%/W. This value is lower than the theoretical efficiency of 5.69 x 10$^{-2}$ \%/W, calculated using the following formulas\cite{Boyd2003}:
\begin{align}
\zeta &= \left( \frac{16 \pi^2 d_{\text{eff}}^2 L P}{\varepsilon_0 c n_1 n_2 \lambda^3} \right)^{1/2} \cdot \text{sinc}^2\left( \frac{\Delta k L}{2} \right) \tag{1} \\
\eta &= \frac{\tanh^2(\zeta)}{\text{sech}^2(\zeta)} = \sinh^2(\zeta) \tag{2}
\end{align}

The dimensionless parameter $\zeta$ defines the strength of the nonlinear SHG interaction. $d_{\text{eff}}$ is the effective nonlinear coefficient of PP-LBGO, $L$ is the length of the nonlinear crystal, and $P$ is the power of the pump beam. The refractive indices of the crystal at fundamental and second-harmonic wavelengths are $n_1$ and $n_2$, respectively. Lastly, $\lambda$ is the fundamental wavelength, $\Delta k$ is the momentum mismatch, and $\eta$ is the SHG efficiency.

This discrepancy likely arises from a combination of imperfect beam focusing or loss during free-space to fiber coupling. We originally intended to use two LBGO crystals to generate photon pairs, using the first crystal to generate UV SHG and the second crystal to subsequently generate SPDC. However, catastrophic laser damage was observed at 3~W. For this reason, a BBO crystal was used for SHG instead, providing 3.45~mW of 266~nm light when pumped with 18~W.

\begin{figure}[htbp]
  \centering
  \begin{subfigure}[t]{0.48\textwidth}
    \centering
    \includegraphics[width=\linewidth]{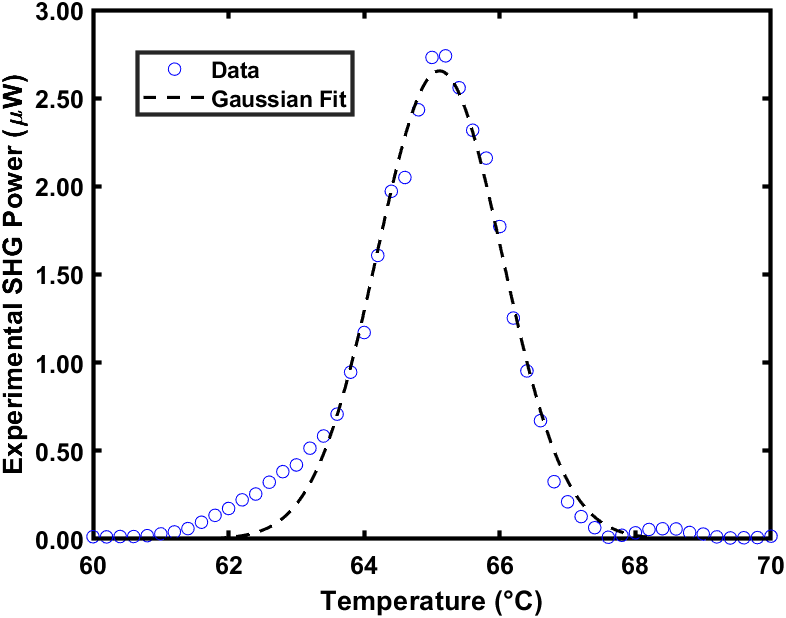}
    \label{fig:setup}
  \end{subfigure}
  \hfill
  \begin{subfigure}[t]{0.48\textwidth}
    \centering
    \includegraphics[width=\linewidth]{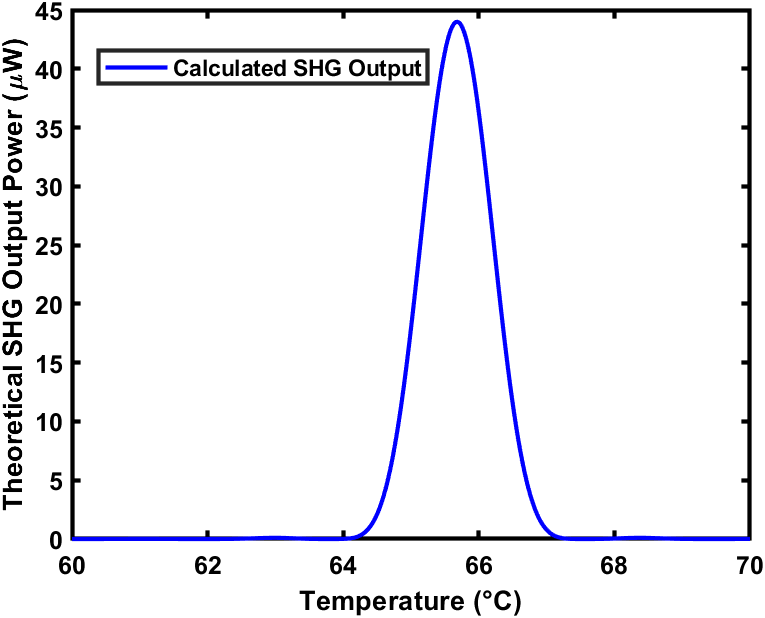}
    \label{fig:spectrum}
  \end{subfigure}
  \caption{Temperature characterization of the generated 266 nm second harmonic signal (left) and theoretical phase matching curve (right). Blue data points are experimentally measured values with the black dashed curve depicting a Gaussian fit of the data. }
  \label{fig:shg_dual}
\end{figure}

SPDC emitted at angles of up to 167.4 mrad are collected by the lens after the chip. This is calculated by forming a right triangle with its height being half of the lens diameter and the length being the distance to L4. The reflective collimator (RC) has a max fiber numerical aperture (NA) of .19 corresponding to a maximum acceptance half angle of 191.2 mrad. In theory, the entirety of the degenerate collinear SPDC should be collected by the fiber collimator. As we discuss later, the actual collection efficiency is estimated to be 16\%.

\begin{figure}[!htbp]
  \centering
  \includegraphics[width=8cm]{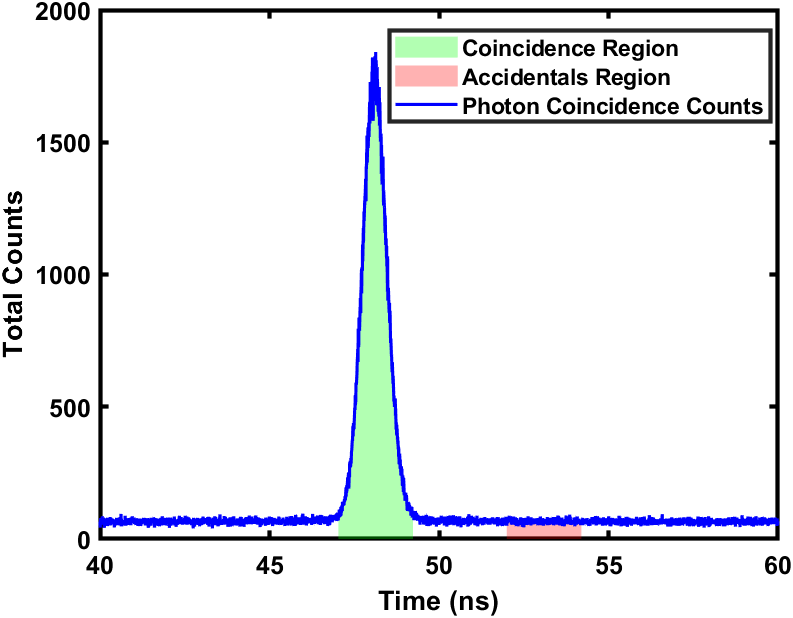}
  \caption{LBGO SPDC Coincidence Histogram Raw Data and Fitted Data. The red highlighted region is used to calculate accidental coincidences in the system. Data is taken at 66.7\,°C where the counts in each SPAD are found to be the highest. The coincidence window is 2.2 ns and is depicted by the green highlighted region.}
  \label{fig:histogram}
\end{figure}

\FloatBarrier

\begin{figure}[!htbp]
  \centering
  \begin{subfigure}[t]{0.48\textwidth}
    \centering
    \includegraphics[width=\linewidth]{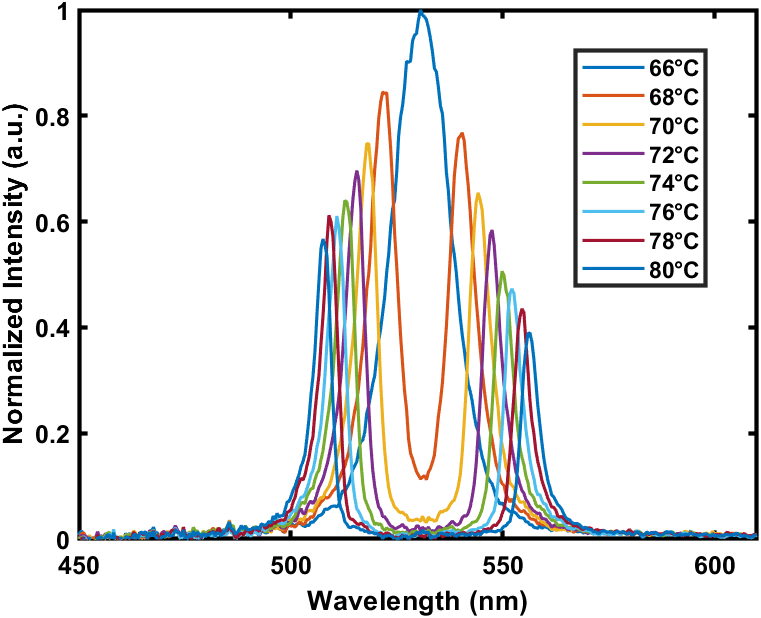}
    \label{fig:fig5a}
  \end{subfigure}
  \hfill
  \begin{subfigure}[t]{0.48\textwidth}
    \centering
    \includegraphics[width=\linewidth]{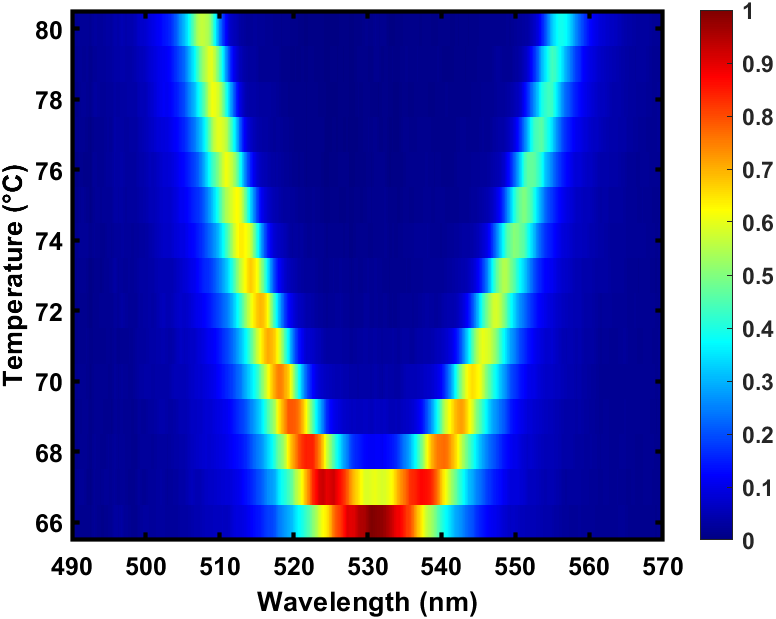}
    \label{fig:fig5b}
  \end{subfigure}
  \caption{LBGO SPDC Spectral Output. (Left) SPDC emission at even temperatures showing tunability from 66 to 80\,°C. (Right) Emission heatmap for visualization across the entire scanned range.}
  \label{fig:spdc_spectrum}
\end{figure}

SPDC is first sent to an EMICCD spectrometer. The spectrum is measured at different temperature points, shown in Fig 5. As the temperature increases, the signal and idler are generated at further apart wavelengths and with lower brightness. A chip temperature of 66.7 °C generates degenerate SPDC with a bandwidth of 16 nm (FWHM) and was used for subsequent efficiency measurements.

\begin{figure}[htbp]
\centering\includegraphics[width=8cm]{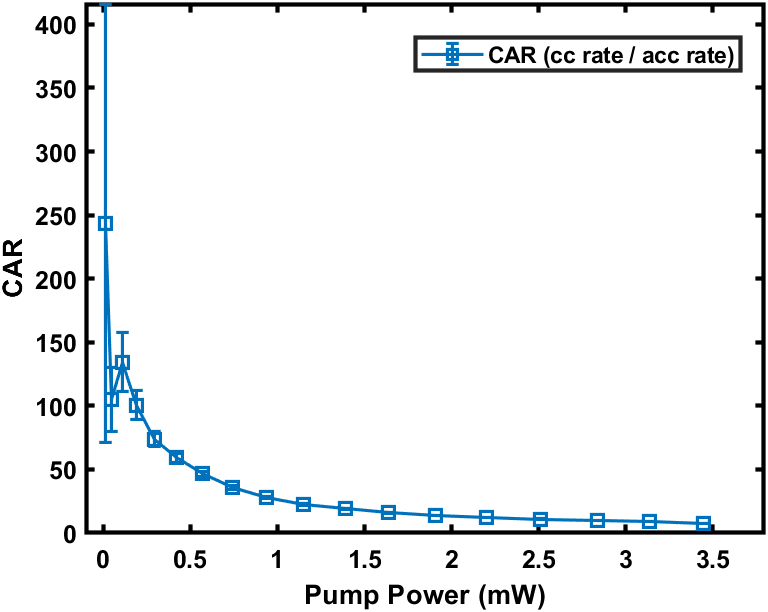}
\caption{CAR at variety of pump powers. The CAR is calculated by calculating the rate within the full 2.2 ns window of the coincidence peak (Fig 4), divided by a region far from the coincidence peak (52 to 54.2 ns) with the same time bin length.}
\end{figure}

To measure the efficiency of the SPDC source, the same collection fiber is used. Pairs are detected in coincidence using the Hanbury-Browne-Twiss setup described previously. Coincidence histograms and singles rates are recorded as the 266 nm pump power is swept. Fig 4 displays one of these histograms, measured at 66.7 °C at max pump power. An electronic delay of 50 nanoseconds (ns) is implemented between the two SPADs, which determines the location of the coincidence peak in the histogram. To ensure all coincidences due to SPDC are measured, a coincidence window of 2.2 ns (larger than the peak FWHM of .800 ns) is used. The accidental rate, caused by uncorrelated SPAD detection events, is measured from 52-54.2 ns. This rate is then subtracted from the measured coincidences, giving a corrected coincidence rate. Figure 6 reports the CAR, which is inversely proportional to the input power. A high value of 243.5 is measured at 11 \textmu W of pump power where the accidental rate is low. A large uncertainty in this maximum value is evident but implies that a value of 416 may be possible with detectors matching the timing resolution of the coincidence counter. A singles-to-pairs ratio of 35:1 and 38:1 is measured at each detector. This non-unity ratio is due to loss and detection inefficiencies. Equations 3-5 give the power scaling of the singles and coincidence rates\cite{Harper2024b}:
\begin{align}
S_1(P) &= 2\,E\,P\,\mu_1\,R\,\eta_1 + \mathrm{Dark}_1 \tag{3}\\
S_2(P) &= 2\,E\,P\,\mu_2\,T\,\eta_1 + \mathrm{Dark}_2 \tag{4}\\
S_{cc}(P) &= 2\,E\,P\,\mu_1\,\mu_2\,R\,T\,\eta_{12} \tag{5}
\end{align}

Here, $S_{1}$, $S_{2}$, and $S_{cc}$ are the singles and coincidence rates as a function of pump power.  $E$ is the device efficiency, $P$ is the pump power at the input of the PP‐LBGO, $\mu_{1}$ and $\mu_{2}$ are the transmission efficiencies of the paths to detectors 1 and 2, respectively, $R$ and $T$ are the beamsplitter reflectance and transmittance, $\eta_{1}$ and $\eta_{12}$ are the average quantum efficiency and joint quantum efficiency, respectively, and $\mathrm{Dark}_{1}$ and $\mathrm{Dark}_{2}$ are the dark count rates of detector 1 and 2. From these equations, a pair generation rate of $1.45\times10^{6}\,\mathrm{pairs\,s^{-1}\,mW^{-1}}$ is implied, along with a collection efficiency of $16\%$.  To determine the collection efficiency, we used the lowest calculated transmission of the two detectors ($18\%$) and accounted for an $87\%$ transmission of the fiber beamsplitter.  The residual loss is likely the result of imperfect collimation and alignment into the fiber collimator.  Table 1 summarizes these results, and the procedure for their calculation is described in a previous work by our group \cite{Harper2024b}.  We have also verified that the singles rate due to sources other than SPDC is below $20\%$ of the singles generated by SPDC.  This upper bound was determined by tuning the 266 nm half‐wave plate (HWP) to pump the crystal with the orthogonal polarization, which is not phase‐matched.  This result suggests that the statistics are not influenced by spurious singles from stray light or fluorescence.

\begin{table}[ht]
\centering
\renewcommand{\arraystretch}{1.1}

\textbf{Table 1. Power Scaling Parameters}

\begin{tabular}{lll}
\hline
\textbf{Parameter} & \textbf{Value} & \textbf{Description} \\ [4pt]
\hline
$m_{1}$   & $(7.77 \pm 0.15)\times 10^{4}\,\mathrm{photons\,s^{-1}\,mW^{-1}}$ & Detector 1 Singles Rate \\[6pt]
$m_{2}$   & $(8.45 \pm 0.19)\times 10^{4}\,\mathrm{photons\,s^{-1}\,mW^{-1}}$ & Detector 2 Singles Rate \\[6pt]
$m_{cc}$  & $(2.25 \pm 0.07)\times 10^{3}\,\mathrm{pairs\,s^{-1}\,mW^{-1}}$   & Coincidence Rate \\[6pt]
$R$       & $51 \pm 2\%$   & Beamsplitter Reflectance \\[6pt]
$T$       & $49 \pm 2\%$   & Beamsplitter Transmittance \\[6pt]
$\eta_{1}$   & $0.30 \pm .04$         & Average QE \\[6pt]
$\eta_{12}$  & $0.09 \pm .01 $         & Average Joint QE \\[6pt]
$E$       & $(1.45 \pm 0.48)\times 10^{6}\,\mathrm{pairs\,s^{-1}\,mW^{-1}}$ & Device Efficiency \\[6pt]
$\mu_{1}$    & $0.18 \pm .04$         & Transmission to Detector 1 \\[6pt]
$\mu_{2}$    & $0.20 \pm .04$         & Transmission to Detector 2 \\[4pt]
\hline
\end{tabular}
\vspace{6pt}
\begin{minipage}{0.85\linewidth}
\footnotesize
Uncertainties of count rates come from standard error of linear fit. Beamsplitter uncertainty is experimentally determined. Detector efficiency is based on manufacturer data, calibrated experimentally, and scaled based on measurements at 532 nm. 
\end{minipage}
\end{table}

\section{Discussion}
We demonstrate SPDC down to 507 nm, but this can be further decreased to the 195 nm transparency of LBGO by shortening the poling period or operating at higher temperatures. The measured pair generation rates of the device are close when compared to current state‐of‐the‐art waveguided PPLN/PPLT QPM devices that report MHz and up to GHz pair‐generation rates. For 532 nm SPDC, BBO is reported to have a PGR of 4700 pairs s$^{-1}$ mW$^{-1}$\cite{Wong2006}, and our PGR of $1.45\times10^{6}$ pairs s$^{-1}$ mW$^{-1}$ is over two orders of magnitude above this value. The pair‐generation rate reported here is currently sufficient to saturate commercially available SPADs at modest pump powers. 

Further, the reported efficiency is higher than that of $\beta$‐BBO while being free of walk‐off effects. As nanofabrication capabilities improve, and the Sellmeier equations of LBGO are better understood, PP‐LBGO could be a viable candidate in both CW UV generation and as a type‐0 SPDC source. Efficiency could further be increased by utilizing a longer crystal, resonating the pump or second harmonic, and via the implementation of waveguides. Waveguiding would be expected to increase efficiencies by fifteen times\cite{Watanabe2019b}. High‐finesse optical parametric oscillator–style designs could also enhance these efficiencies by recycling pump power.

The measured SPDC output of the PP‐LBGO device is expected to enable a variety of new possibilities for spectroscopy. The nondegenerate SPDC output of PP‐LBGO would encompass the entire excitation spectrum of multiple biomarkers, for example, flavin adenine dinucleotide (FAD) and its five redox derivatives\cite{Schwinn2020}. While there is much overlap between the fluorescence spectra of each species, they can be distinguished by their distinct lifetimes. After locking in on a specific lifetime associated with a particular species, the fluorescence intensity of that signal can be measured to obtain information regarding fluorophore concentration. Extending entangled fluorescence to the UV range broadly applies to most fluorophores.

While the present work reports a low SPDC efficiency in comparison to waveguided devices, it demonstrates the viability of LBGO as a material enabling access to the deep UV and UV‐pumped nonlinear processes. The discrepancy between experimental and theoretical SHG efficiency suggests our experimental focusing and collection methods could be improved, and that the utility of LBGO could be potentially high with the further development of precise nanofabrication techniques. Advances in precision of the QPM poling periods and the incorporation of waveguides are expected to dramatically increase SHG and SPDC efficiencies while enabling tunability of the SPDC output.

\section{Conclusion}

In conclusion, we have shown that periodically poled LBGO and quasi‐phase‐matching techniques are feasible for generating CW light in the UV wavelength range with our experimentally measured SHG efficiency being $(3.52 \pm 0.04)\times10^{-3}\,\%\;/\mathrm{W}$. When pumping the periodically poled LBGO with 266\,nm light and analyzing the SPDC output, we observed a maximum CAR of 244 and a maximum pair generation rate of $(1.45 \pm 0.47)\times10^{6}\,\mathrm{pairs\,s^{-1}\,mW^{-1}}$. Periodically poled LBGO is therefore expected to be a promising candidate as a nonlinear material as fabrication and waveguiding methods are improved.

\section*{Funding}
National Science Foundation (DGE-2139433). Department of Energy (DE-SC0022089)

\section*{Acknowledgments}
The authors gratefully acknowledge the critical support and infrastructure provided for this work by Caltech. R.T. was supported by the National Science Foundation Graduate Research Fellowship Program under Grant no. DGE-2139433 and by the Department of Energy under Grant no. DE-SC0022089. Any opinion, findings, and conclusions or recommendations expressed in this material are those of the authors and do not necessarily reflect the views of the National Science Foundation nor the Department of Energy.

\section*{Disclosures}
The authors declare no conflicts of interest.

\section*{Data availability}
Data underlying the results presented in this paper are not publicly available at this time but may be obtained from the authors upon reasonable request.

\end{document}